\documentclass[12pt]{article}

\usepackage{epsfig}

\def\square{\kern1pt\vbox{\hrule height 1.2pt
\hbox{\vrule width 1.2pt\hskip 3pt
\vbox{\vskip 6pt}\hskip 3pt\vrule width 0.6pt}
\hrule height 0.6pt}\kern1pt}
\def\ltwid{\mathrel{\raise.3ex\hbox{$<$\kern-.75em\lower1ex\hbox{$\sim$}}}}
\def\gtwid{\mathrel{\raise.3ex\hbox{$>$\kern-.75em\lower1ex\hbox{$\sim$}}}}

\begin{document}

\begin{titlepage}
\begin{flushright}
CCTP-2011-23 \\ UFIFT-QG-11-08
\end{flushright}

\vspace{0.5cm}

\begin{center}
\bf{Primordial Gravitational Waves Enhancement}
\end{center}

\vspace{0.3cm}

\begin{center}
Maria G. Romania$^{\dagger}$ and  N. C. Tsamis$^{\ddagger}$
\end{center}
\begin{center}
\it{Institute of Theoretical \& Computational Physics, and \\
Department of Physics, University of Crete \\
GR-710 03 Heraklion, HELLAS.}
\end{center}

\vspace{0.2cm}

\begin{center}
R. P. Woodard$^{\ast}$
\end{center}
\begin{center}
\it{Department of Physics, University of Florida \\
Gainesville, FL 32611, UNITED STATES.}
\end{center}

\vspace{0.3cm}

\begin{center}
ABSTRACT
\end{center}
\hspace{0.3cm} We reconsider the enhancement of primordial 
gravitational waves that arises from a quantum gravitational 
model of inflation. A distinctive feature of this model is 
that the end of inflation witnesses a brief phase during which
the Hubble parameter oscillates in sign, changing the usual 
Hubble friction to anti-friction. An earlier analysis of this 
model was based on numerically evolving the graviton mode 
functions after guessing their initial conditions near the 
end of inflation. The current study is based on an equation 
which directly evolves the normalized square of the magnitude. 
We are also able to make a very reliable estimate for the 
initial condition using a rapidly converging expansion for 
the sub-horizon regime. Results are obtained for the energy 
density per logarithmic wave number as a fraction of the critical
density. These results exhibit how the enhanced signal depends 
upon the number of oscillatory periods; they also show the 
resonant effects associated with particular wave numbers.

\vspace{0.3cm}

\begin{flushleft}
PACS numbers: 04.30.-m, 04.62.+v, 98.80.Cq
\end{flushleft}

\vspace{0.1cm}

\begin{flushleft}
$^{\dagger}$ e-mail: romania@physics.uoc.gr \\
$^{\ddagger}$ e-mail: tsamis@physics.uoc.gr \\
$^{\ast}$ e-mail: woodard@phys.ufl.edu
\end{flushleft}

\end{titlepage}

\section{Introduction}

The case for a phase of accelerated expansion {\it (inflation)}
during the very early universe is strong. One reason is that
we can observe widely separated parts of the early universe
which seem to be in thermal equilibrium with one another
\cite{infl}. If one assumes the universe never underwent a
period of inflation, there would not have been time for this
thermal equilibrium to be established by causal processes.
Without primordial inflation the number of causally distinct
regions in our past light-cone at the time of recombination
is over $10^3$, and it would be $10^9$ at the time of
nucleosynthesis.

There is no strong indication for what caused primordial
inflation. A natural mechanism for inflation can be found 
within gravitation -- which, after all, plays the dominant 
role in shaping cosmological evolution -- by supposing that 
the bare cosmological constant $\Lambda$ is not unnaturally 
small but rather large and positive.
\footnote{Here ``large'' means a $\Lambda$ induced by a
matter scale which can be as high as $10^{18} \, GeV$. Then,
the value of the dimensionless coupling constant can be as
high as $\, G \Lambda \sim 10^{-4}$ rather than the putative 
value of $10^{-122}$.}
Because $\Lambda$ is constant in {\it space}, no special
initial condition is needed to start inflation. We also
dispense with the need to employ a new, otherwise undetected
scalar field. However, $\Lambda$ is constant in {\it time}
as well, and classical physics can offer no natural mechanism
for stopping inflation once it has begun \cite{stability}.
Quantum physics can: accelerated expansion continually rips
virtual infrared gravitons out of the vacuum \cite{gravitons}
and these gravitons attract one another, thereby slowing
inflation \cite{NctRpw1}. This is a very weak effect for
$\, G \Lambda \ll 1$, but a cumulative one, so inflation 
lasts a long time for no other reason than that gravity 
is a weak interaction \cite{NctRpw1}.

This screening mechanism may be clear enough on the 
perturbative level but it has two frustrating features.
The first is that, because inflationary particle production
is a 1-loop effect, the gravitational response to it is 
delayed until 2-loop order. The second frustration is that 
the 2-loop effect becomes unreliable just when it starts 
to get interesting. The effective coupling constant is 
$\, G \Lambda H t \,$ and higher loops are insignificant 
as long as it is small. But {\it all} loops become comparable 
when $\, G \Lambda H t \,$ becomes of order one, and the 
correct conclusion then is that perturbation theory breaks 
down. The breakdown occurs not because any single 
graviton-graviton interaction gets strong but rather 
because there are so many of them.

We believe it may be possible to derive a non-perturbative
resummation technique by extending the stochastic method which
Starobinsky devised for the same purpose in scalar potential
models \cite{AAS,NctRpw2,sqed}. However, generalizing this 
technique to gravity is a difficult problem \cite{Miao}. This 
paper is part of an effort which is based on the idea of 
{\it guessing} the most cosmologically significant part of 
the effective field equations of quantum gravity. While there 
is no chance of guessing the full effective field equations, 
it might be possible to guess just enough to correctly describe 
the evolution of the scale factor $a(t)$ for a homogeneous and 
isotropic geometry, using what we know from perturbation theory 
about how the back-reaction effect scales. Such simple cosmological 
models were recently constructed \cite{NctRpw3,NctRpw4} and are 
reviewed in Section 2. Some basics of linearized gravitons needed
in this work are the subject of Section 3. The possibility of 
getting an enhancement of high frequency gravity waves within 
this class of models was first investigated in \cite{romania}. 
In this paper we re-investigate this possibility using more 
accurate calculational techniques and we present our results 
in a way more appropriate for the needs of gravitational wave 
experiments. In Section 4 we review the enhancement mechanism 
while in Section 5 we derive an equation for the square of the 
magnitude of the mode functions and describe our improved 
evolution strategy. Our results are presented in Section 6. 
Their physical consequences and our concluding remarks comprise 
Sections 7 and 8.

\section{The Cosmological Model}

In a previous paper \cite{NctRpw3} we proposed a phenomenological 
model which can provide evolution beyond perturbation theory. 
In one sentence, we constructed an {\it effective} conserved 
stress-energy tensor $T_{\mu\nu} [g]$ which modifies the 
gravitational equations of motion:
\footnote{Hellenic indices take on spacetime values while Latin 
indices take on space values. Our metric tensor $g_{\mu\nu}$ has 
spacelike signature and our curvature tensor equals:
$R^{\alpha}_{~\beta\mu\nu} \equiv 
\Gamma^{\alpha}_{~\nu\beta, \mu} +
\Gamma^{\alpha}_{~\mu\rho} \;
\Gamma^{\rho}_{~\nu\beta} -
(\mu \leftrightarrow \nu)$.
The initial Hubble parameter is $3H^2_0 \equiv \Lambda$.}
\begin{equation}
G_{\mu\nu} \; \equiv \;
R_{\mu\nu} \, - \, \frac12 g_{\mu\nu} \, R \; = \;
- \Lambda \, g_{\mu\nu} \, + \,
8 \pi G \, T_{\mu\nu}[g] 
\;\; . \label{eom1}
\end{equation}
and which, we hope, contains the most cosmologically
significant part of the full effective quantum
gravitational equations.

What form to guess for $T_{\mu\nu}[g]$ was motivated by
what we seek to do, and by what we know from perturbation
theory. We seek to describe cosmology, which implies
homogeneous and isotropic geometries. When specialized to
such a geometry the full effective stress tensor must take
the perfect fluid form and we lose nothing by assuming
that generally:
\begin{equation}
T_{\mu\nu}[g] \; = \;
(\rho + p) \, u_{\mu} \, u_{\nu} \, + \,
p \, g_{\mu\nu}
\;\; . \label{Tmn}
\end{equation}
The relation between $p[g]$, $\rho[g]$ and $u_{\mu}[g]$ is
heavily constrained by stress-energy conservation, but it is
possible to specify one function for free. It turns out to be
computationally simplest to take this free function to be the
pressure \cite{NctRpw3}. We further require the pressure to be
an ordinary function of some non-local scalar which grows like
the number of e-foldings when specialized to de Sitter. If the 
pressure is to grow the way we know it does from perturbation 
theory \cite{sqed}, and to eventually end inflation, then a 
simple choice has the form \cite{NctRpw3}:
\begin{equation}
p[g](x) \; = \;
\Lambda^2 \, f[- G \Lambda \, X](x) 
\qquad , \qquad
X \, \equiv \, \frac{1}{\square} \, R
\;\; , \label{pressure}
\end{equation}
where the function $f$ grows without bound and satisfies:
\begin{equation}
f[- G \Lambda \, X] \; = \;
- G \Lambda \, X \, + \, 
O\Big[ (G \Lambda)^2 \Big]
\;\; , \label{fincr}
\end{equation}
and where the scalar d'Alembertian:
\begin{equation}
\square \, \equiv \;
\frac{1}{\sqrt{-g}} \;
\partial_{\mu} \Big( \,
g^{\mu\nu} \sqrt{-g} \; \partial_{\nu} \, \Big)
\;\; , \label{box}
\end{equation}
is defined with retarded boundary conditions.
The induced energy density $\rho[g]$ and 4-velocity
$u_{\mu}[g]$ are determined, up to their initial
value data, from stress-energy conservation:
\begin{equation}
D^{\mu} \, T_{\mu\nu} \; = \; 0
\;\; . \label{cons1}
\end{equation}
The 4-velocity was chosen to be timelike and
normalized:
\begin{equation}
g^{\mu\nu} \, u_{\mu} u_{\nu} = -1
\qquad \Longrightarrow \qquad
u^{\mu} \, u_{\mu ; \nu} = 0
\;\; . \label{u}
\end{equation}

The homogeneous and isotropic evolution
\footnote{The line element in co-moving coordinates is
$ds^2 = -dt^2 + a^2(t) \, d{\vec x} \cdot d{\vec x}$.
In terms of the scale factor $a$, the Hubble parameter
equals $H(t) = {\dot a} \, a^{-1}$ and the deceleration
parameter equals $q(t) = - a \, {\ddot a} \, {\dot a}^{-2} 
= -1 - {\dot H} \, H^{-2} \equiv -1 + \epsilon(t)$.}
of this model -- using a combination of numerical
and analytical methods -- revealed the following
basic features:
\footnote{In \cite{NctRpw3}, our analytical results
were obtained for any function $f$ satisfying
(\ref{fincr}) and growing without bound, our numerical
results for the choice: $f(x) = \exp(x) - 1$.}
\\
$\bullet \,$ After the onset and during the era of 
inflation, the source $X(t)$ grows while the curvature 
scalar $R(t)$ and Hubble parameter $H(t)$ decrease. 
\\ [5pt]
$\bullet \,$ Inflationary evolution dominates roughly 
until we reach a critical point $X_{cr}$ defined by:
\begin{equation}
1 - 8 \pi G \Lambda \, f[ - G \Lambda \, X_{cr}] 
\; \equiv \; 0
\;\; . \label{Xcr}
\end{equation}
$\bullet \,$ The epoch of inflation ends close to but 
before the universe evolves to the critical time. This 
is most directly seen from the deceleration parameter
since initially $q(t=0) = -1$ while at criticality
$q(t=t_{cr}) = +\frac12$. 
\\ [5pt]
$\bullet \,$ Oscillations in $R(t)$ become significant 
as we approach the end of inflation; they are centered
around $R = 0$, their frequency equals:
\begin{equation}
\omega \; = \; 
G \Lambda H_0 \sqrt{72 \pi \, f_{cr}'}
\;\; , \label{omega}
\end{equation}
where $H_0$ is the constant inflationary Hubble parameter,
and their envelope is linearly falling with time.
\\ [5pt]
$\bullet \,$ During the oscillations era, although 
there is net expansion, the oscillations of $H(t)$ 
take it to small negative values for short time 
intervals -- a feature conducive to rapid reheating; 
those of ${\dot H}(t)$ take it to positive values 
for about half the time; and, those of $a(t)$ are 
centered around a linear increase with time.

A novel feature of this class of models is the
existence of an oscillatory regime of short duration
which commences towards the very end of the inflationary
era. During this period ${\dot H}(t)$ is positive about 
half the time, which represents a violation of the weak 
energy condition. Such a violation cannot occur in 
classically stable theories \cite{Cline} but it can be 
driven by quantum effects of the type we seek to model 
without endangering stability \cite{woodard1}.

\section{Linearized Gravitons}

In terms of the full metric field $g_{ij}(x) \,$, 
the fluctuating graviton field $h_{ij}^{TT}(x)$ is 
defined as: 
\begin{equation}
g_{ij}(t, {\bf x}) \; = \;
a^2(t) \Big[ \delta_{ij} + 
\sqrt{32 \pi G \,} \, h_{ij}^{TT}(t, {\bf x}) \Big]
\;\; . \label{hij}
\end{equation}
The free field expansion of the graviton field is:
\begin{equation}
h_{ij}^{TT}(t, {\bf x}) \; = \;
\int \frac{d^3 k}{(2 \pi)^3} \sum_{\lambda}
\left\{ u(t, k) \, e^{i {\bf k} \cdot {\bf x}} 
\epsilon_{ij}({\bf k}, \lambda) \, 
\alpha({\bf k}, \lambda) \, + \, (c.c.) \right\}
\;\; , \label{hijexp}
\end{equation}
where $(c.c.)$ denotes complex conjugation, the 
polarizations $\epsilon_{ij}({\bf k}, \lambda)$
and operators $\alpha({\bf k}, \lambda)$ obey:
\begin{eqnarray}
\epsilon_{ij}({\bf k}, \lambda) \;
\epsilon_{ij}^*({\bf k}, \lambda')
&\!\! = \!\!&
\delta_{\lambda \lambda'}
\quad , \quad
\epsilon_{ii}({\bf k}, \lambda) 
\, = \,
k_i \, \epsilon_{ij}({\bf k}, \lambda)
\, = \, 0
\;\; , \qquad \label{polarizations} \\
\Big[ \, \alpha({\bf k}, \lambda) \, , \,
\alpha^{\dagger}({\bf k}', \lambda') \, \Big]
&\!\! = \!\!&
\delta_{\lambda \lambda'} \, (2 \pi)^3 \,
\delta^3 ({\bf k} - {\bf k'})
\; \; , \label{operators}
\end{eqnarray}
and the mode functions $u(t, k)$ satisfy:
\begin{equation}
{\ddot u}(t, k) \, + \,
3 H(t) \, {\dot u}(t, k) \, + \,
\frac{k^2}{a^2(t)} \, u(t, k) 
\; = \; 0
\;\; , \label{modeseqn}
\end{equation}
with the Wronskian associated with the two 
solutions of (\ref{modeseqn}) equaling:
\begin{equation}
u \, {\dot u}^* \, - \,
{\dot u} \, u^* \; = \;
i a^{-3}
\;\; . \label{wronskian}
\end{equation}

We shall be interested in the energy $E(t, k)$ at 
time $t$ of a mode with wavenumber $k$. The simplest
way to derive this is to exploit the fact that the
physical degrees of freedom of linearized gravitons
have the same dynamics with those of a massless, 
minimally coupled scalar field $\varphi(x)$.
\footnote{The analogous computation within the
linearized graviton theory should only make an 
$O(1)$ change to the result.}
The scalar field Lagrangian density is:
\begin{equation}
{\cal L}(x) \; = \;
- \frac12 \, \sqrt{-g} \, g^{\mu\nu} \,
{\partial}_{\mu} \varphi \,
{\partial}_{\nu} \varphi 
\; = \;
\frac12 \, a^3(t) \, {\dot \varphi}^2 \, - \,
\frac12 \, \nabla \varphi \cdot \nabla \varphi
\;\; . \label{Lmmcs} 
\end{equation}
The Langangian diagonalizes in momentum space:
\begin{equation}
L(t) = 
\int d^3 x \; {\cal L}(x) = 
\int \frac{d^3 k}{(2 \pi)^3} 
\left\{ \frac12 \, a^3(t) \, 
\Big\vert \dot{\widetilde{\varphi}}(t, {\bf k}) \Big\vert^2
- \, \frac12 \, a(t) \, k^2 \,
\Big\vert \widetilde{\varphi}(t, {\bf k}) \Big\vert^2
\right\}
\label{Ldiag}
\end{equation}
so that any mode with wavenumber ${\bf k}$ evolves 
independently as a harmonic oscillator $q(t)$ with
time-dependent mass $m(t) \equiv a^3(t)$ and angular
frequency $\omega(t) \equiv k \, a^{-1}(t)$:
\begin{eqnarray}
q(t) &\!\! = \!\!& 
u(t, k) \, A \, + \, u^*(t, k) \, A^{\dagger}
\quad , \quad
\Big[ \, A \, , \, A^{\dagger} \, \Big] \, = \, 1
\;\; , \label{SHO1} \\
E_{SHO}(t) &\!\! = \!\!&
\frac12 \, a^3(t) \, {\dot q}^2(t) \, + \,
\frac12 \, a(t) \, k^2 \, q(t)
\;\; . \label{SHO2}
\end{eqnarray}

At any instant $t$ the minimum energy is $E_{\rm min}(t, k) 
= \frac12 k \, a^{-1}(t)$. However since both the mass
and angular frequency are time-dependent, the state with
minimum energy at one time instant is not the state with
minimum energy at another time instant; there is particle
production as time evolves. The Bunch-Davies vacuum $\vert
\Omega \rangle$ is the minimum energy state in the distant
past and the expectation value of the energy operator 
(\ref{SHO2}) in its presence equals: 
\begin{equation}
\langle \Omega \vert \, E(t, k) \, \vert \Omega \rangle 
\; = \;
\frac12 \, a^3(t) \, \vert {\dot u}(t, k) \vert^2 \, + \,
\frac12 \, k^2 a(t) \, \vert u(t, k) \vert^2
\;\; . \label{E}
\end{equation} 
A fair measure of the excess energy $\Delta E(t, k)$ 
acquired during time evolution in any one wavenumber
is obtained by subtracting the instantaneous minimum
energy from (\ref{E}):
\begin{eqnarray}
\Delta E(t, k) &\!\! \equiv \!\!& 
\langle \Omega \vert \, E(t, k) \, \vert \Omega \rangle 
\, - \, E_{\rm min}(t, k)
\nonumber \\
&\!\! = \!\!&
\frac12 \, a^3(t) \, \vert {\dot u}(t, k) \vert^2 \, + \,
\frac12 \, k^2 a(t) \, \vert u(t, k) \vert^2 \, - \,
\frac{k}{2 a}
\;\; . \label{DeltaE}
\end{eqnarray}

\section{The Enhancement Mechanism}

The oscillatory phase is a very distinctive feature
of these models and in \cite{romania} we investigated
the possibility of gravitational wave enhancement due
to its presence. There are two very plausible physical
arguments that convinced us this is a worthwhile inquiry:
\\ [5pt]
- During the oscillations era the Hubble parameter $H(t)$
changes sign and this, in turn, changes the sign of the 
``friction'' term $3 H {\dot u}$ in the evolution equation 
(\ref{modeseqn}) obeyed by the mode functions $u(t, k)$. 
For $H(t) > 0$ this term tends to reduce 
$\vert {\dot u}(t, k) \vert$ whereas it tends to increase
$\vert {\dot u}(t, k) \vert$ when $H(t) < 0$. What happens
to the magnitude $\vert u(t, k) \vert$ depends upon where
$u(t, k)$ is in its own oscillations when $H(t)$ changes
sign but the change from ``friction'' to ``anti-friction''
can clearly strengthen the amplitude in some cases.
\\ [5pt]
- The oscillations era is characterized by the frequency
$\omega$ given by (\ref{omega}). Gravitational waves of
frequency close to $\omega$ can resonate and their amplitude
can increase.

The first effort to evolve (\ref{modeseqn}) through the 
oscillatory phase was done in \cite{romania}. As expected,
it is the {\it near-horizon} modes that experience enhancement:
the natural time scale of their $u(t, k)$ is close to the 
inverse of the oscillatory frequency $\omega$ and we get
a significant resonance response. 
\footnote{Here and throughout, {\it super-, sub-, near- 
horizon} is with respect to the modes $k_{\rm cr}$ whose 
first horizon crossing occured at criticality, when the 
transition from the inflationary to the oscillating era 
occured: $k_{\rm cr} = H(t_{\rm cr}) \, a(t_{\rm cr})$.}
When converted to current frequencies, the main conclusion
of \cite{romania} is the enhancement of gravitational waves 
with frequencies somewhat less than $10^{10} \, H\!z$.
In obtaining these results, however, certain assumptions
were necessary since we do not possess exact forms for 
the two linearly independent solutions of (\ref{modeseqn}) 
during the oscillatory regime. Nor do we know which linear
combination of these two solutions is the actual mode
function as we do not know the linear combination 
coefficients.
\footnote{The actual mode function is the coefficient
of the annihilation operator in the free field expansion 
of the graviton.}
The latter are determined by knowledge of the initial
conditions at criticality. Because the post-inflationary
scale factor effectively describes an overall linear 
expansion on which the oscillations are superimposed 
\cite{NctRpw3}, in \cite{romania} we solved (\ref{modeseqn})
for a linearly expanding $a(t)$ -- which can be done
exactly -- and then numerically superimposed the effect
of the oscillations. We also had to make an ``educated
guess'' regarding the initial conditions at criticality.

In re-visiting the subject, we have developed a method
-- to be described in the next Section -- which is 
considerably more accurate and, therefore, leads to 
robust conclusions.

\section{The Evolution Strategy}

$\bullet \;$ {\it The Variable $M(t, k)$}
\\ 
We wish to derive an equation for the quantity $M(t, k)$:
\begin{equation}
M(t, k) \equiv u(t, k) \, u^*(t, k) 
= \vert u(t, k) \vert^2
\;\; , \label{M}
\end{equation}
because it is directly related to the tensor power 
spectrum $\Delta^2_h(t, k)$:
\begin{eqnarray}
\Delta^2_h(t, k) &\!\! \equiv \!\!&
32 \pi G \, \frac{k^3}{2 \pi^2} \int
d^3x \; e^{-i {\bf k} \cdot {\bf x}} \;
\langle \Omega \, \vert \, 
h_{ij}^{TT}(t, {\bf x}) \; h_{ij}^{TT}(t, {\bf 0})
\, \vert \, \Omega \rangle
\nonumber \\
&\!\! = \!\!&
\frac{16 \pi G}{\pi} \, k^3 \, M(t, k)
\;\; . \label{Deltah}
\end{eqnarray}

From the definition of $M$ it follows that:
\begin{eqnarray}
{\dot M} &\!\! = \!\!& 
{\dot u} \, u^* \, + \, u \, {\dot u}^*
\;\; , \label{Mdot} \\
{\ddot M} &\!\! = \!\!&
{\ddot u} \, u^* \, + \, 2 {\dot u} \, {\dot u}^*
\, + \, u \, {\ddot u}^*
\;\; . \label{Mddot}
\end{eqnarray}
By using the fact that ${\ddot u}$ satisfies
(\ref{modeseqn}) we conclude:
\begin{equation}
{\ddot M} \, + \, 3 H {\dot M} \, + \,
\frac{2 k^2}{a^2} \, M 
\; = \; 
2 {\dot u} \, {\dot u}^*
\;\; . \label{Meqn0}
\end{equation}
By subtracting the square of (\ref{wronskian}) from that 
of (\ref{Mdot}), we can express the right hand side of 
(\ref{Meqn0}) in terms of $M$ and ${\dot M}$:
\begin{equation}
{\dot u} \, {\dot u}^* \; = \;
\frac{1}{4 M} \, \left[ 
{\dot M}^2 \, + \, \frac{1}{a^6} \right]
\;\; , \label{udotu}
\end{equation}
and obtain the desired equation: 
\begin{equation}
{\ddot M} \, + \, 3 H {\dot M} \, + \,
\frac{2 k^2}{a^2} \, M 
\; = \; 
\frac{1}{2 M} \left[
{\dot M}^2 \, + \, \frac{1}{a^6} \right]
\;\; . \label{Meqn}
\end{equation}
The goal is to find $M(t, k)$ such that (\ref{Meqn}) is
obeyed. An exact solution is beyond our abilities but we 
can divide the full time evolution range into separate
intervals and obtain reliable approximate expressions 
for $M(t, k)$ within each of these.
\\ [5pt]
$\bullet \;$ {\it The Evolution of $M(t, k)$: Inflation}
\\
- During the inflationary era, it makes sense to adopt 
a scheme that works accurately for any kind of mode and,
at the same time, avoids numerical evolution for as long
as possible. A method that seems optimal is the development 
of an asymptotic series expansion for $M(t, k)$ in powers 
of $H^2 a^2 \div k^2$:
\begin{equation}
M(t, k) \; = \;
\frac{1}{2k \, a^2} \, \Bigg\{
1 \, + \, \alpha(t) \left( \frac{H a}{k} \right)^2 \, + \,
\beta(t) \left( \frac{H a}{k} \right)^4 \, + \, \dots
\Bigg\}
\;\; . \label{Mexp}
\end{equation}
Substituting the above in (\ref{Meqn}) allows us 
to determine the leading coefficients $\alpha(t), 
\, \beta(t)$ of the series. The final form for the 
asymptotic expansion of $M(t, k)$ becomes:
\begin{eqnarray}
M(t, k) &\!\! = \!\!&
\frac{1}{2k \, a^2} \, \Bigg\{
1 \, + \,
\Big( 1 - \frac{\epsilon}{2} \Big) 
\left( \frac{H a}{k} \right)^2 \, + \,
\Bigg[ \frac94 \, \epsilon \, - \,
\frac{21}{8} \, \epsilon^2 \, + \,
\frac34 \, \epsilon^3 \, + \,
\nonumber \\
& \mbox{} &
\hspace{2.3cm}
+ \, \Big( \, \frac74 - \frac{3 \epsilon}{4} \, \Big) 
\frac{\dot{\epsilon}}{H} \, + \,
\frac{\ddot{\epsilon}}{8 H^2} \Bigg]
\left( \frac{H a}{k} \right)^4 \, + \,
\dots \Biggr\}
\;\; . \label{Mexp2}
\end{eqnarray}

As long as $\epsilon$ does not get large, the series
(\ref{Mexp2}) converges rapidly and we can use it to
evolve all the way to within, say, 2 e-foldings before
first horizon crossing.
\footnote{The error is in ignoring terms proportional to 
$\left( \frac{H a}{k} \right)^6$ and higher. Even when we 
reach 2 e-foldings before first horizon crossing that is 
very small: 
$\left( \frac{H a}{k} = e^{-2} \right)^6 = e^{-12}$.}
We shall, therefore, adopt this method and evolve very
accurately: 
(i) any {\it sub-horizon} mode all the way to criticality, 
(ii) any {\it near-horizon} mode until, say, 2 e-foldings 
before criticality, and 
(iii) any {\it super-horizon} mode until, say, 2 e-foldings 
before first horizon crossing. 
Afterwards, in all cases equation (\ref{Meqn}) is evolved 
numerically.

The important cosmological parameters in the inflationary 
era are \cite{NctRpw3,romania}:
\begin{eqnarray}
a(t) &\!\! = \!\!&
a_{\rm cr} \, e^{-N}
\;\; , \label{ainfl} \\
H(t) &\!\! \simeq \!\!&
\frac13 \, \omega \, \sqrt{ 4N + \frac43}
\;\; , \label{Hinfl} \\
{\dot H}(t) &\!\! \simeq \!\!&
- \frac{2 H^2}{4N + \frac43}
\;\; , \label{Hdotinfl} \\
\epsilon(t) &\!\! \simeq \!\!&
\frac{2}{4N + \frac43}
\;\; , \label{epsiloninfl}
\end{eqnarray}
where $N$ is the number of e-foldings before criticality.
The initial conditions used in the numerical analysis are 
those inherited from (\ref{Mexp2}) at the appropriate time.
\\ [5pt]
$\bullet \;$ {\it The Evolution of $M(t, k)$: Oscillations}
\\
During the oscillatory era the important cosmological 
parameters are \cite{NctRpw3,romania}:
\begin{eqnarray}
a(t) &\!\! = \!\!&
a_{\rm cr} \, C_2 \Big[ \,
C_1 \, + \, \omega \, \Delta t \, + \,
{\sqrt 2} \, \cos( \omega \, \Delta t + \phi ) \Big]
\;\; , \label{aosc} \\
H(t) &\!\! = \!\!&
\frac{ \omega \Big[ \, 1 - 
{\sqrt 2} \, \sin( \omega \, \Delta t + \phi ) \Big] }
{ C_1 \, + \, \omega \, \Delta t \, + \,
{\sqrt 2} \, \cos( \omega \, \Delta t + \phi ) }
\;\; , \label{Hosc} \\
{\dot H}(t) &\!\! = \!\!&
- H^2(t) \, - \,
\frac{ \omega^2 {\sqrt 2} \, \cos( \omega \, \Delta t + \phi ) }
{ C_1 \, + \, \omega \, \Delta t \, + \,
{\sqrt 2} \, \cos( \omega \, \Delta t + \phi ) }
\;\; , \label{Hdotosc} \\
\epsilon(t) &\!\! = \!\!&
1 \, + \,
\frac{ {\sqrt 2} \, \cos( \omega \, \Delta t + \phi )
\Big[ \, C_1 \, + \, \omega \, \Delta t \, + \,
{\sqrt 2} \, \cos( \omega \, \Delta t + \phi ) \Big] }
{ \Big[ \, 1 - {\sqrt 2} \, \sin( \omega \, \Delta t + \phi ) 
\Big]^2 }
\;\; , \qquad \label{epsilonosc}
\end{eqnarray}
where $\, \Delta t \equiv t - t_{\rm cr} \,$ measures 
time with respect to criticality. In this regime we 
analyze equation (\ref{Meqn}) numerically. The parameters 
$( \phi, \, C_1, \, C_2 )$ in (\ref{aosc}-\ref{epsilonosc}) 
are chosen to match the outcome from the inflationary epoch 
(\ref{ainfl}-\ref{epsiloninfl}) at criticality, where $N=0$ 
and $\Delta t = 0$:
\begin{eqnarray} 
\phi &\!\! = \!\!&
\arcsin \! \left( 
\frac {\sqrt{2} - \sqrt{2970}}{56} \right)
\; \approx \; 
- \frac{\pi}{2}
\;\; , \label{phi} \\
C_1 &\!\! = \!\!&
\frac{\sqrt{27}}{2} - 
\frac{\sqrt{27}}{2} \, \sin\phi -
\sqrt{2} \, \cos\phi
\; \approx \; 3
\;\; , \label{C1} \\
C_2 &\!\! = \!\!&
\frac{1}{C_1 + \sqrt{2} \, \cos\phi}
\; \approx \; \frac16
\;\; . \label{C2}
\end{eqnarray}
\\
$\bullet \;$ {\it An Observable}
\\
To connect with physical measurements, consider the 
excess energy $\Delta E(t, k)$ at time $t$ of a mode 
with wavenumber $k$. It is given by equation 
(\ref{DeltaE}) or, equivalently, by:
\begin{equation}
\Delta E(t, k) \; = \;
\frac{a^3 {\dot M}^2}{8 M} \, + \,
\Big[ 2 k \, a^2 M - 1 \Big]^2
\;\; , \label{DeltaE2}
\end{equation}
where we have used (\ref{udotu}, \ref{M}). We shall be 
interested in any excess energy $\Delta E$ acquired 
during time evolution through the oscillating regime.
The resulting excess energy density $\Delta \rho$ is:
\begin{equation}
\Delta \rho (t, k) \; = \;
\int \frac{d^3 k}{[ \, 2 \pi \, a(t) \, ]^3} \; 
\Delta E(t, k)
\; = \;
\frac{1}{2 \pi^2 a^3(t)}
\int dk \; k^2 \; \Delta E(t, k)
\;\; . \label{Deltarho} 
\end{equation}

Perhaps of more relevance for gravity wave detectors
is the amount of gravitational waves energy density
$\Delta \rho$ per wavenumber $k$, and divided by the 
critical density $\rho_{\rm cr}$:
\begin{eqnarray}
\frac{d \;\;\;}{d \ln k} \, \Omega_{\rm gw} (t, k)
&\!\! \equiv \!\!&
\frac{1}{\rho_{\rm cr}} \;
\frac{d \;\;\;}{d \ln k} \, \Delta \rho (t, k)
\label{detectors1} \\
&\!\! = \!\!&
\frac{4 G \, k^3}{3 \pi \, H^2_{\rm now} \, a^3(t)} \;
\Delta E(t, k)
\;\; . \label{detectors2}
\end{eqnarray}

\section{The Results}

We first define dimensionless variables:
\begin{equation}
\tau \; \equiv \; \omega \, \Delta t
\;\; , \;\;
\kappa \; \equiv \; \frac{k}{\omega a_{\rm cr}}
\;\; , \;\;
\alpha \; \equiv \; \frac{a}{a_{\rm cr}}
\;\; , \;\;
{\cal H} \; \equiv \; \frac{H}{\omega}
\;\; , \;\;
{\cal M} \; \equiv \; 2 k \, a^2 \, M 
\label{dimless}
\end{equation}
and re-express in terms of them the evolution equation
(\ref{Meqn}):
\begin{eqnarray}
\frac{d^2 \cal{M}}{d \tau^2} \, + \,
{\cal H} \, \frac{d \cal{M}}{d \tau} \, + \,
{\cal H}^2 \, (4-2 \epsilon) \, {\cal M} \, + \,
\frac{2 \kappa^2}{\alpha^2} \, 
\left[ \frac{1}{\cal{M}} \, - \, {\cal M} \right]
\; = \; 
\frac{1}{2 {\cal M}} 
\left( \frac{d {\cal M}}{d \tau} \right)^2 
\label{Meqndimless}
\end{eqnarray}
the asymptotic series expansion (\ref{Mexp}):
\begin{eqnarray}
{\cal M}(t, k) &\!\! = \!\!&
1 \, + \,
\Big( 1 - \frac{\epsilon}{2} \Big) 
\left( \frac{{\cal H} \alpha}{\kappa} \right)^2 
\label{Mexpdimless} \\
& \mbox{} &
\hspace{-1.3cm}
+ \, \Bigg[ \frac94 \, \epsilon \, - \,
\frac{21}{8} \, \epsilon^2 \, + \,
\frac34 \, \epsilon^3 \, + \,
\Big( \, \frac74 - \frac{3 \epsilon}{4} \, \Big) 
\frac{\dot{\epsilon}}{H}
+ \, \frac{\ddot{\epsilon}}{8 H^2} \Bigg]
\left( \frac{{\cal H} \alpha}{\kappa} \right)^4 \, + \,
\dots 
\;\; , \nonumber
\end{eqnarray}
as well as the excess energy (\ref{DeltaE2}):
\begin{equation}
\Delta E \; = \;
\omega \left\{
\frac{\alpha {\cal H}^2}{4 \kappa \, \cal{M}} \,
\Big({\cal M} - \frac{1}{2 {\cal H}} 
\frac{\partial \cal{M}}{\partial \tau} \Big)^2
\, + \,
\frac{\kappa}{4 \alpha \cal{M}}
\Big[ {\cal M} - 1 \Big]^2
\right\}
\;\; , \label{DeltaEdimless} \\
\end{equation}
and the observable (\ref{detectors2}):
\begin{eqnarray}
\frac{d \;\;\;}{d \ln k} \, \Omega_{\rm gw} 
& \!\! = \!\! &
G \omega^2 \times
\frac{4}{3 \pi} \;
\frac{1}{{\cal H}^2} \;
\Big( \frac{\kappa}{\alpha} \Big)^4 
\nonumber \\
& \mbox{} &
\times \, \frac{1}{4 \cal{M}} \left\{
\left( \frac{{\cal H} \alpha}{\kappa} \right)^2 
\Big[ {\cal M} \, - \, \frac{1}{2 {\cal H}}
\frac{\partial \cal{M}}{\partial \tau} \Big]^2
\, + \, 
\Big[ {\cal M} - 1 \Big]^2 \right\}
\;\; . \label{detectorsdimless}
\end{eqnarray}

We then discretize the time interval:
\begin{equation}
\tau \; \rightarrow \; 
\tau_i \, \equiv \, i \, \Delta \tau
\;\; , \label{tau_i}
\end{equation}
and the resulting discretized evolution equation
(\ref{Meqndimless}) determines ${\cal M}_{i+2}$ 
in terms of ${\cal M}_{i+1}$ and ${\cal M}_i$: 
\begin{eqnarray}
{\cal M}_{i+2} &\!\! = \!\!&
2 {\cal M}_{i+1} \, - \, {\cal M}_i \, - \,
{\cal H}_i \, \Delta\tau \,
( {\cal M}_{i+1} - {\cal M}_i ) \, + \,
{\cal H}^2 \, \Delta\tau^2 (4 - 2\epsilon) \, {\cal M}_i
\nonumber \\
& \mbox{} &
+ \, \frac{( {\cal M}_{i+1} - {\cal M}_i )^2}{2 {\cal M}_i}
\, + \,
\frac{2 \kappa^2 \Delta\tau^2}{\alpha_i^2}
\left(\frac{1}{{\cal M}_i} - {\cal M}_i
\right)
\;\; . \label{M_ieqn}
\end{eqnarray}
The results of the combined evolution, using the 
asymptotic series (\ref{Mexpdimless}) until either 
2 e-foldings before first horizon crossing or 
criticality
\footnote{For {\it super-horizon} or 
{\it near/sub-horizon} modes, respectively.}
-- whichever comes first -- and numerical integration 
of (\ref{M_ieqn}) thereafter, are presented in Figures 
1-14. It is important to note the following: \\
- The conditions used for initializing the numerical
integration at 2 e-foldings before first horizon
crossing are provided by evaluating (\ref{Mexpdimless}) 
at this point. \\
- The conditions used for initializing the numerical
integration at criticality are provided by matching
the inflationary solution (\ref{ainfl}-\ref{epsiloninfl})
with the oscillatory solution(\ref{aosc}-\ref{epsilonosc})
at this point, so that the three parameters $(\phi, \,
C_1, \, C_2)$ take on the values (\ref{phi}-\ref{C2}). \\
- The dimensionless wavenumber $\kappa_{\rm cr}$ which 
underwent first horizon crossing at $t = t_{\rm cr}$ 
and is the wavenumber differentiating {\it super-horizon} 
from {\it sub-horizon} modes equals:
\begin{equation}
\kappa_{\rm cr} \, = \,
{\cal H}_{\rm cr} \, \alpha_{\rm cr}
\, = \,
\frac{2}{\sqrt{27} \,} \, \approx \, 0.38
\;\; . \label{kappacr}
\end{equation}
In creating Figures 1-11 we have chosen $136$ values 
of wavenumbers $\kappa$ ranging from $0.05$ 
{\it (super-horizon)} to $10$ {\it (sub-horizon)}. \\
- Inspection of Figure 14 makes evident the wild time
dependence of the observable $\; {\cal H}^2 \times 
(\frac{d \;\;\;}{d \ln k} \, \Omega_{\rm gw}) \,$.
Note that had we not multiplied the observable by 
${\cal H}^2$ even wilder variations would occur when 
${\cal H}$ passes through zero. Hence it is not clear 
to identify where the transition from the oscillatory 
to the radiation domination era took place. A reasonable 
{\it assumption} is that the transition occured when 
${\cal H} > 0$ and $\epsilon = 2$. This determines 
the corresponding time $\tau$ to equal:
\begin{equation}
radiation
\quad \Longrightarrow \quad
{\cal H} > 0 \;\; \& \;\; \epsilon = 2 
\quad \Longrightarrow \quad
\tau \, \approx \, 2 \pi \, N_{\rm osc}
\;\; . \label{taurad}
\end{equation}
- To study the effect of the duration of the oscillatory
regime on the enhancement, we have displayed the results 
for $N_{\rm osc}$ number of oscillation periods within 
the regime. The values of $N_{\rm osc}$ used range from 
$N_{\rm osc} = 0$ -- which corresponds to the usual 
transition from inflation to radiation domination -- 
to $N_{\rm osc} = 10$. Figures 1-11 present the results 
-- for ${\cal H} > 0$ and $\epsilon = 2$ -- in ascending 
order of $N_{\rm osc}$ values. \\
- The dimensionless factor $G \omega^2$ is {\it not} 
included in the evaluation of $\, (\frac{d \;\;\;}{d \ln k} 
\, \Omega_{\rm gw}) \,$ as seen in Figures 1-11, 14.

\begin{figure}
\centerline{\epsfig{file=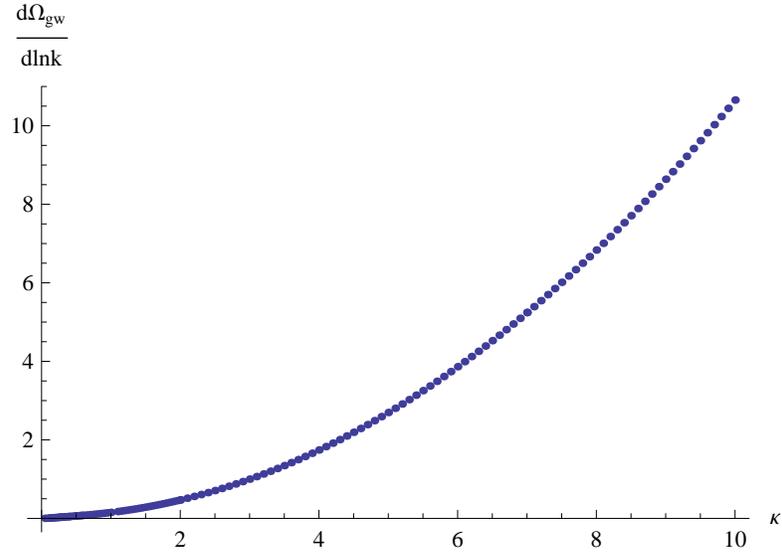,height=2.9in}}
\caption{\footnotesize Fraction of the energy density per 
wavenumber divided by the critical density
\break \mbox{} \hspace{1.75cm}
in gravity waves from our signal, for zero periods of 
oscillations.}
\label{rho0}
\end{figure}

\begin{figure}
\centerline{\epsfig{file=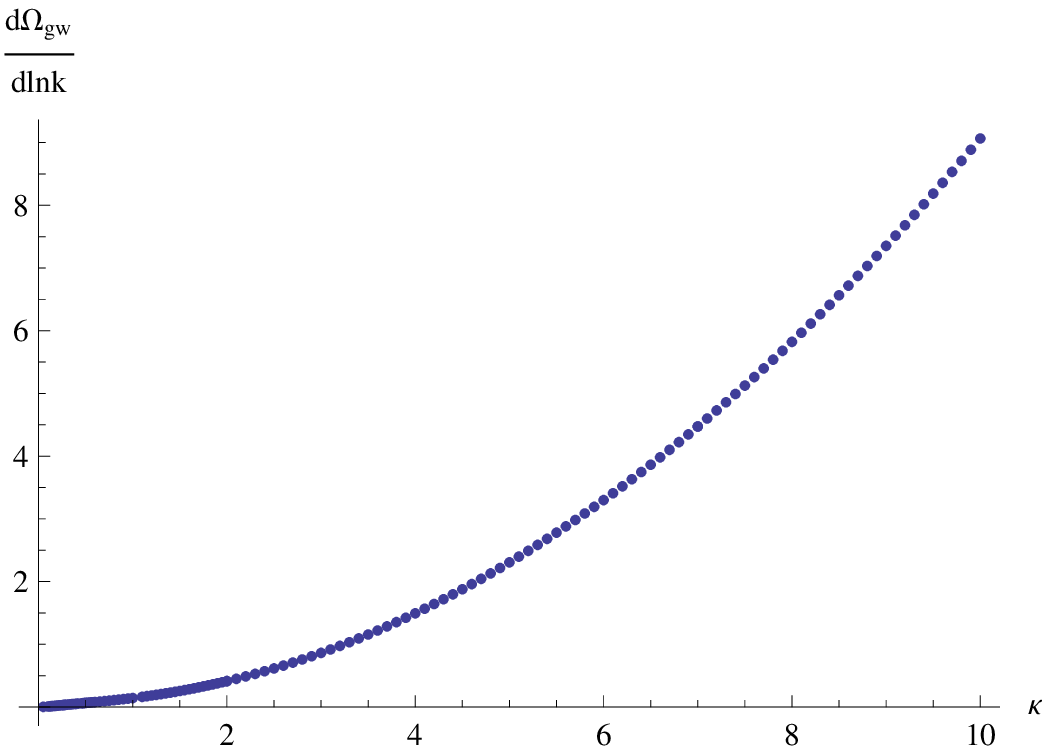,height=2.9in}}
\caption{\footnotesize Fraction of the energy density per 
wavenumber divided by the critical density
\break \mbox{} \hspace{1.75cm}
in gravity waves from our signal, at the beginning of the 
first oscillation.}
\label{rho1}
\end{figure}

\begin{figure}
\centerline{\epsfig{file=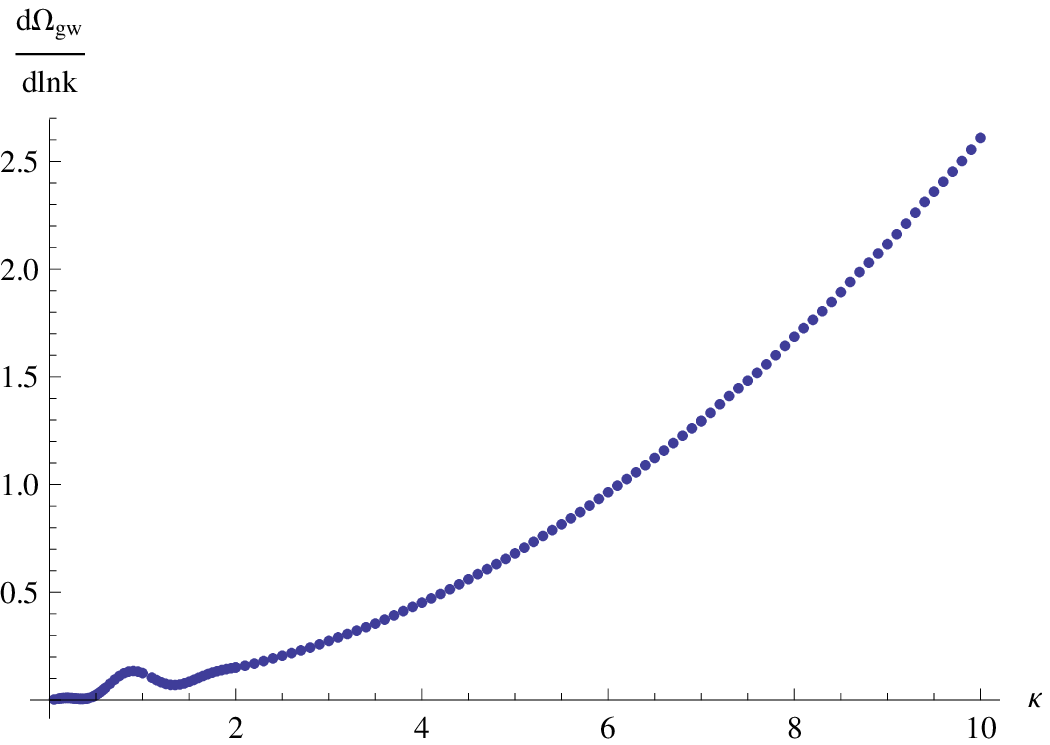,height=2.9in}}
\caption{\footnotesize Fraction of the energy density per 
wavenumber divided by the critical density
\break \mbox{} \hspace{1.75cm}
in gravity waves from our signal, at the beginning of the 
second oscillation.}
\label{rho2}
\end{figure}

\begin{figure}
\centerline{\epsfig{file=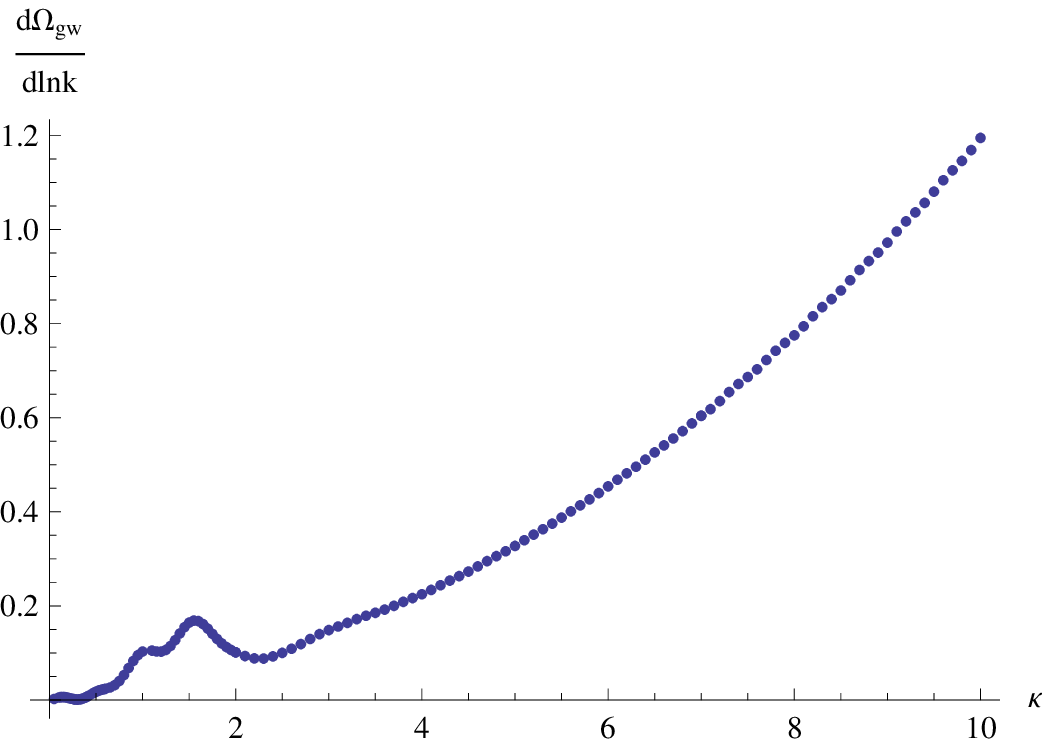,height=2.9in}}
\caption{\footnotesize Fraction of the energy density per 
wavenumber divided by the critical density
\break \mbox{} \hspace{1.75cm}
in gravity waves from our signal, at the beginning of the
third oscillation.}
\label{rho3}
\end{figure}

\begin{figure}
\centerline{\epsfig{file=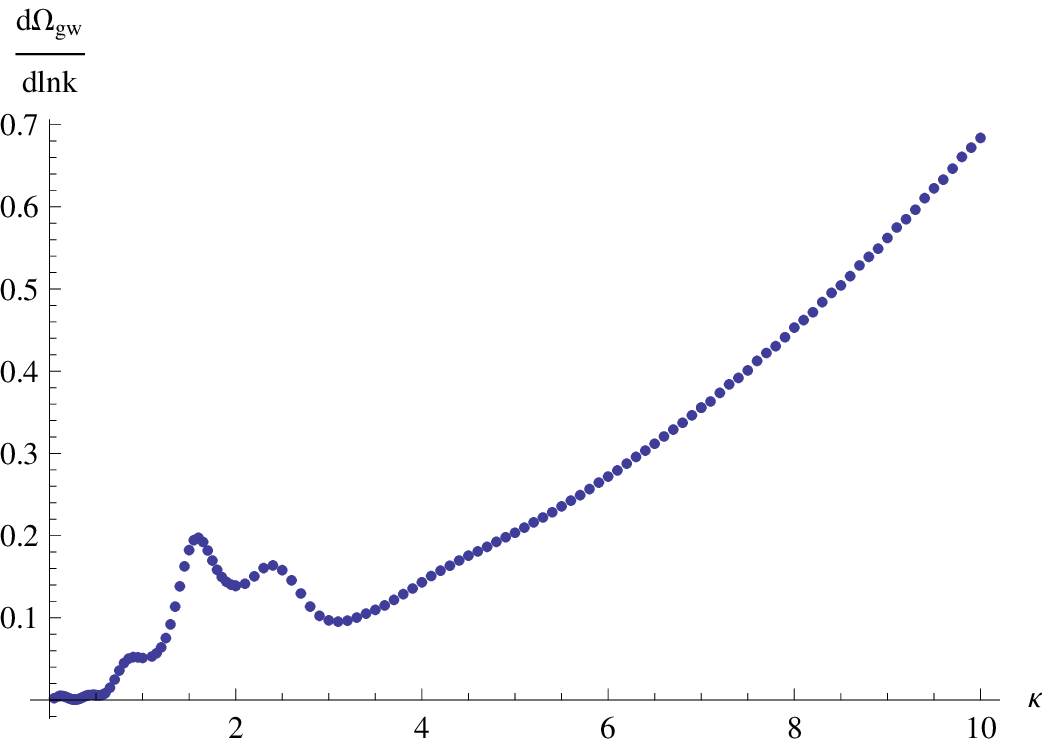,height=2.9in}}
\caption{\footnotesize Fraction of the energy density per 
wavenumber divided by the critical density
\break \mbox{} \hspace{1.75cm}
in gravity waves from our signal, at the beginning of the
fourth oscillation.}
\label{rho4}
\end{figure}

\begin{figure}
\centerline{\epsfig{file=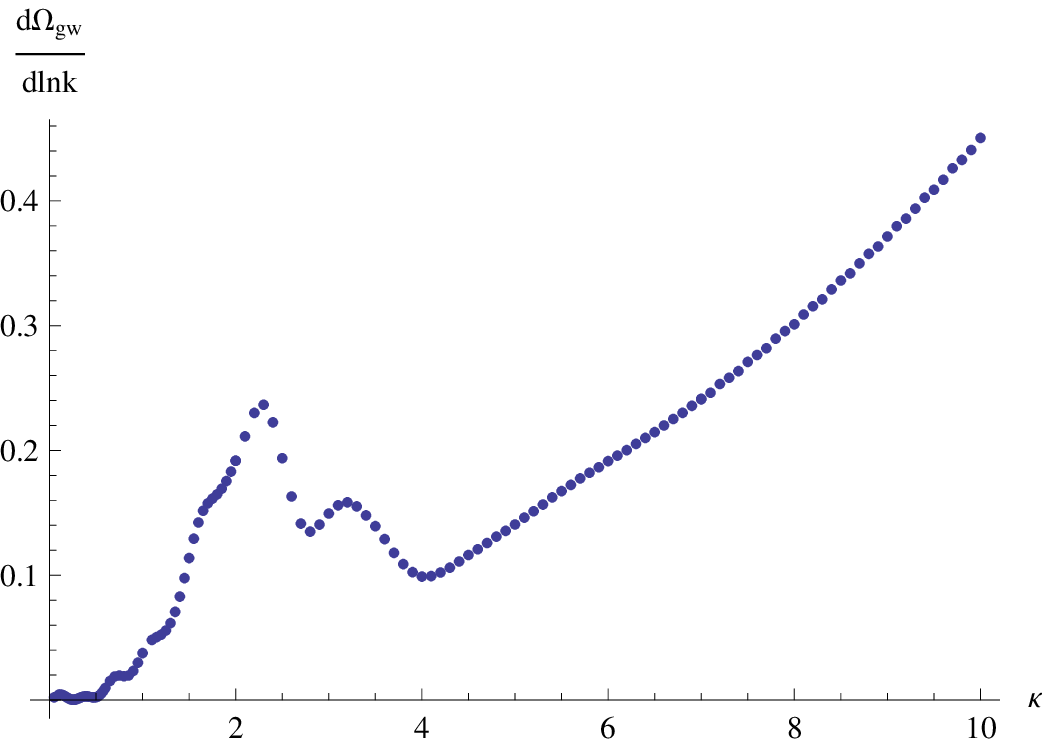,height=2.9in}}
\caption{\footnotesize Fraction of the energy density per 
wavenumber divided by the critical density
\break \mbox{} \hspace{1.75cm}
in gravity waves from our signal, at the beginnng of the
fifth oscillation.}
\label{rho5}
\end{figure}

\begin{figure}
\centerline{\epsfig{file=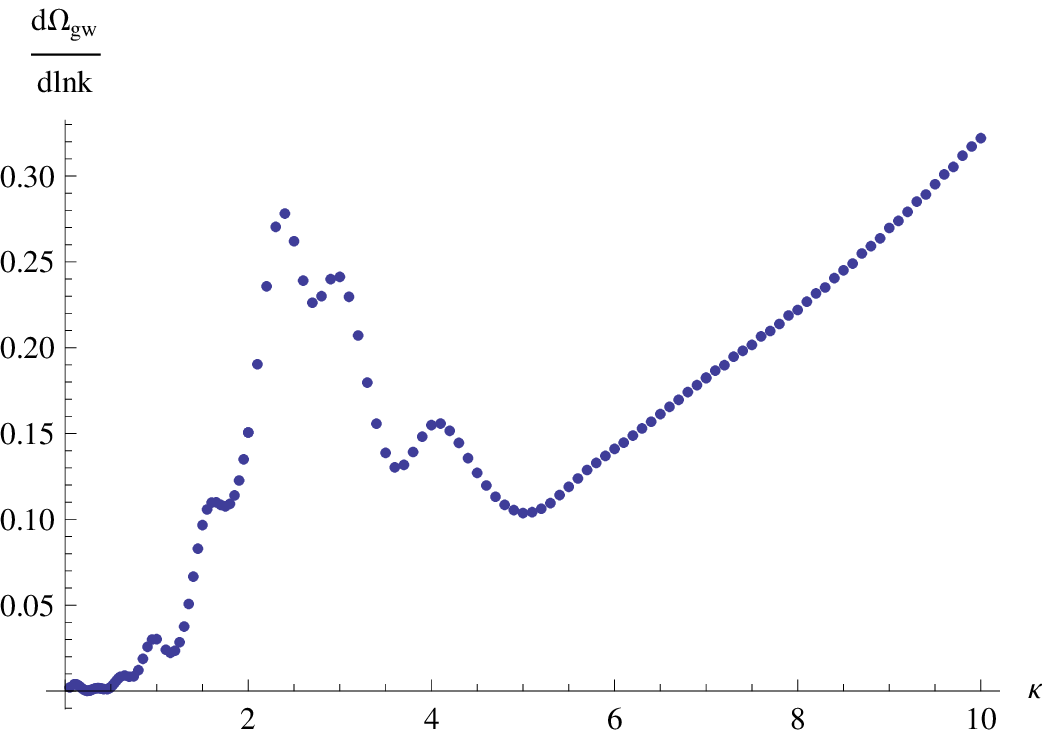,height=2.9in}}
\caption{\footnotesize Fraction of the energy density per 
wavenumber divided by the critical density
\break \mbox{} \hspace{1.75cm}
in gravity waves from our signal, at the beginning of the
sixth oscillation.}
\label{rho6}
\end{figure}

\begin{figure}
\centerline{\epsfig{file=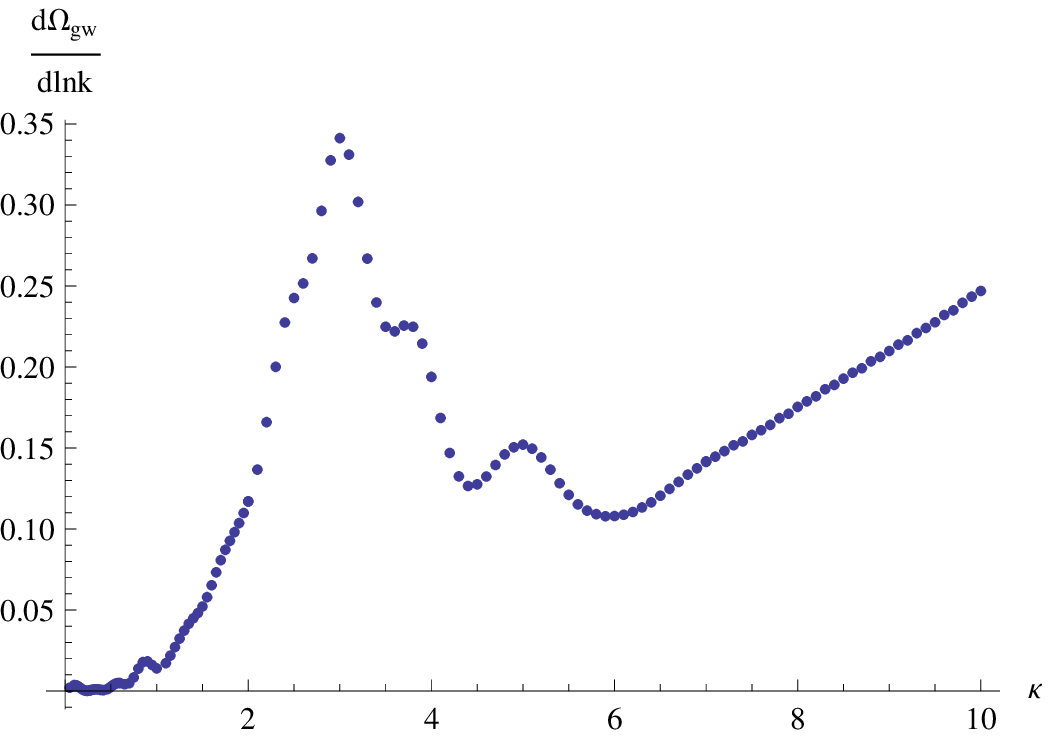,height=2.9in}}
\caption{\footnotesize Fraction of the energy density per 
wavenumber divided by the critical density
\break \mbox{} \hspace{1.75cm}
in gravity waves from our signal, at the beginning of the
seventh oscillation.}
\label{rho7}
\end{figure}

\begin{figure}
\centerline{\epsfig{file=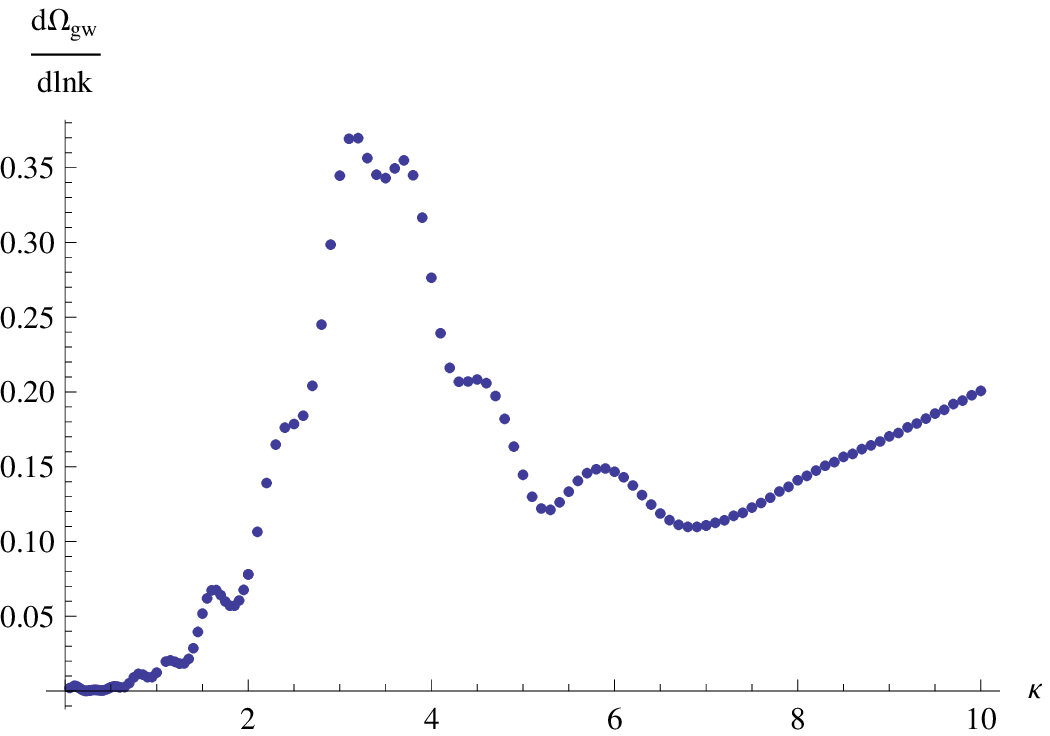,height=2.9in}}
\caption{\footnotesize Fraction of the energy density per 
wavenumber divided by the critical density
\break \mbox{} \hspace{1.75cm}
in gravity waves from our signal, at the beginning of the
eighth oscillation.}
\label{rho8}
\end{figure}

\begin{figure}
\centerline{\epsfig{file=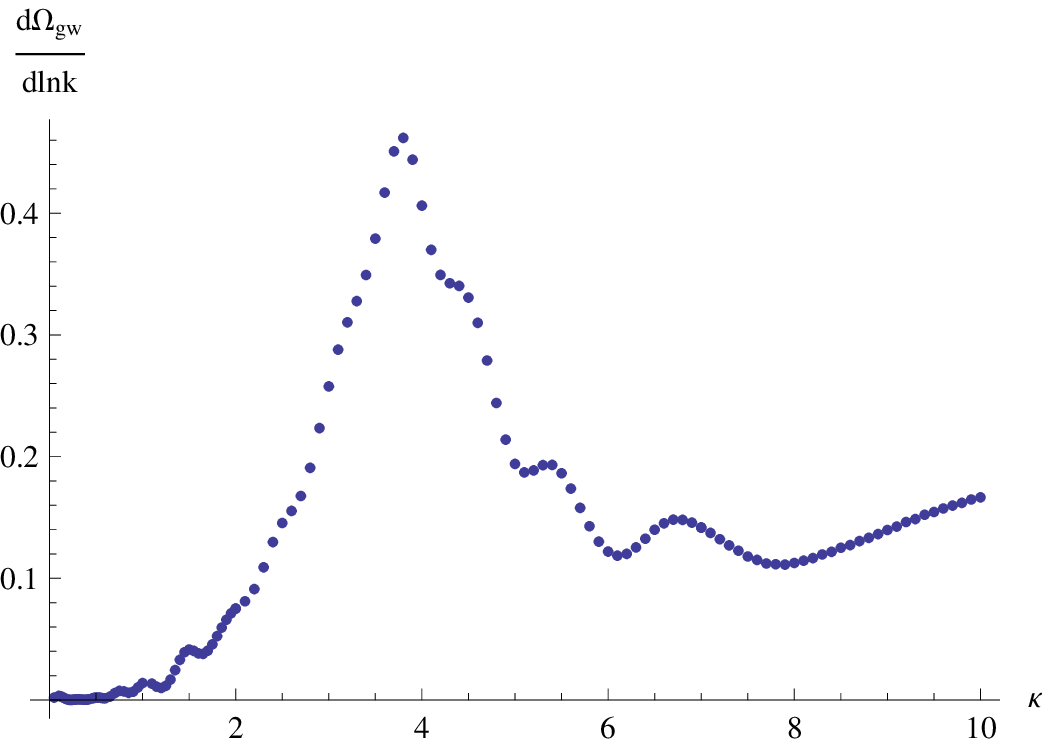,height=2.9in}}
\caption{\footnotesize Fraction of the energy density per 
wavenumber divided by the critical 
\break \mbox{} \hspace{2.10cm}
density in gravity waves from our signal, at beginning of
ninth oscillation.}
\label{rho9}
\end{figure}

\begin{figure}
\centerline{\epsfig{file=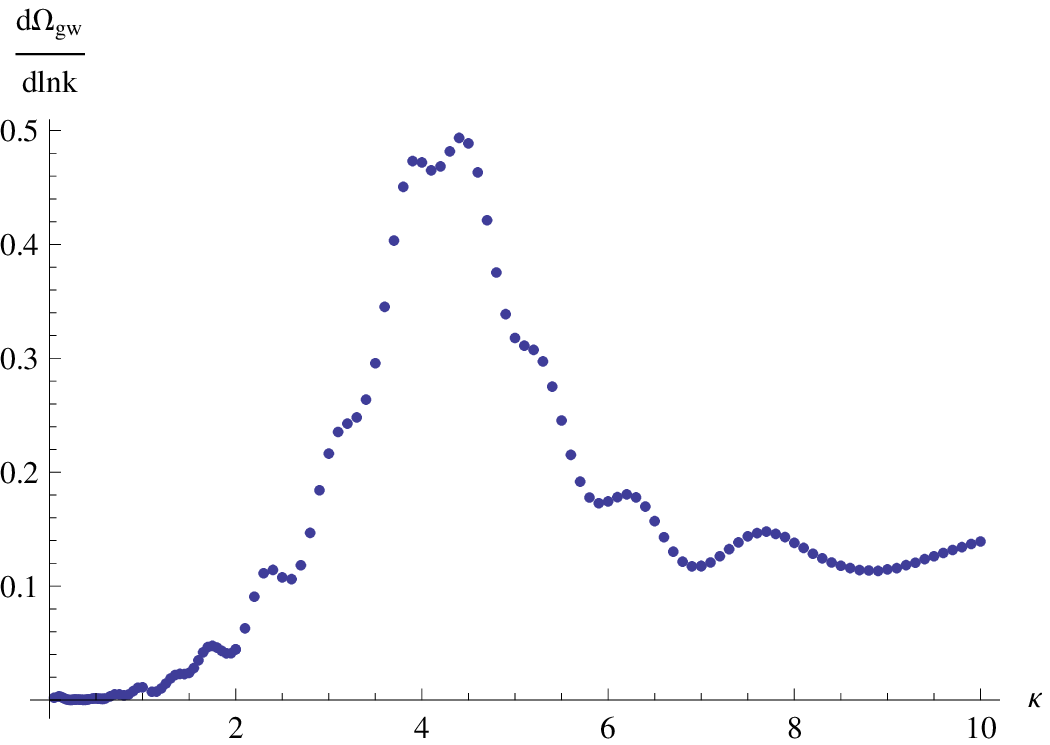,height=2.9in}}
\caption{\footnotesize Fraction of the energy density per 
wavenumber divided by the critical 
\break \mbox{} \hspace{2.10cm}
density in gravity waves from our signal, at beginning 
of tenth oscillation.}
\label{rho10}
\end{figure}

\begin{figure}
\centerline{\epsfig{file=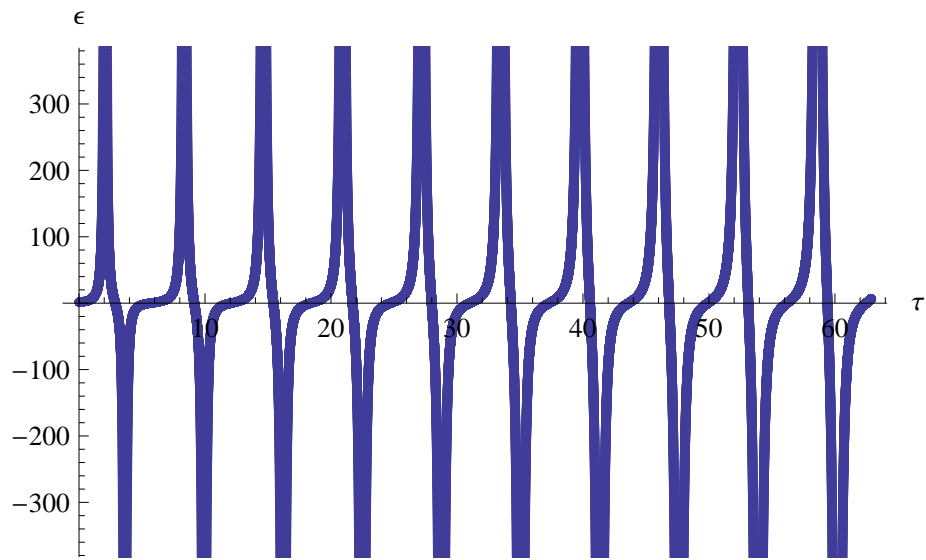,height=2.9in}}
\caption{\footnotesize Time evolution of parameter 
$\; \epsilon \equiv - {\dot H} \, H^{-2} \,$ during 
the oscillatory regime.}
\label{epsilon}
\end{figure}

\begin{figure}
\centerline{\epsfig{file=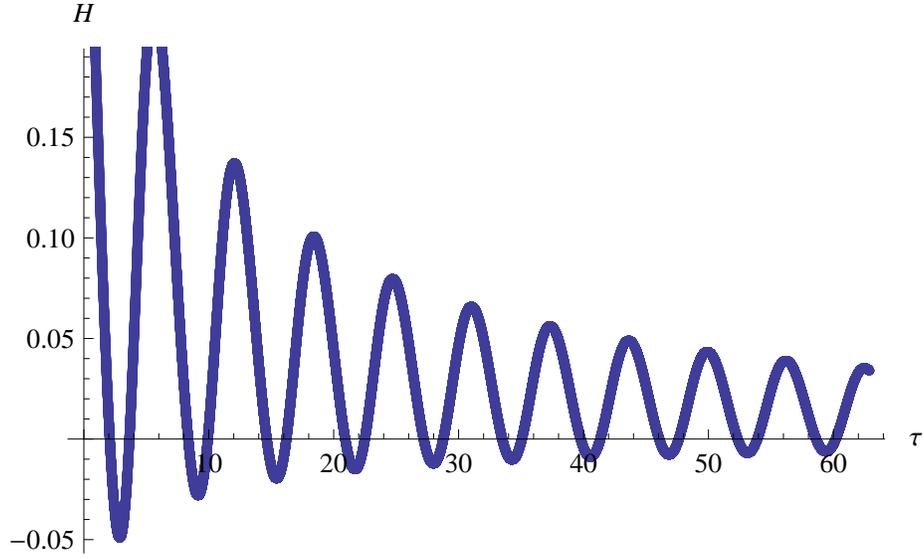,height=2.9in}}
\caption{\footnotesize Time evolution of the Hubble 
parameter ${\cal H}$ during the oscillatory regime.}
\label{hubble}
\end{figure}

\begin{figure}
\centerline{\epsfig{file=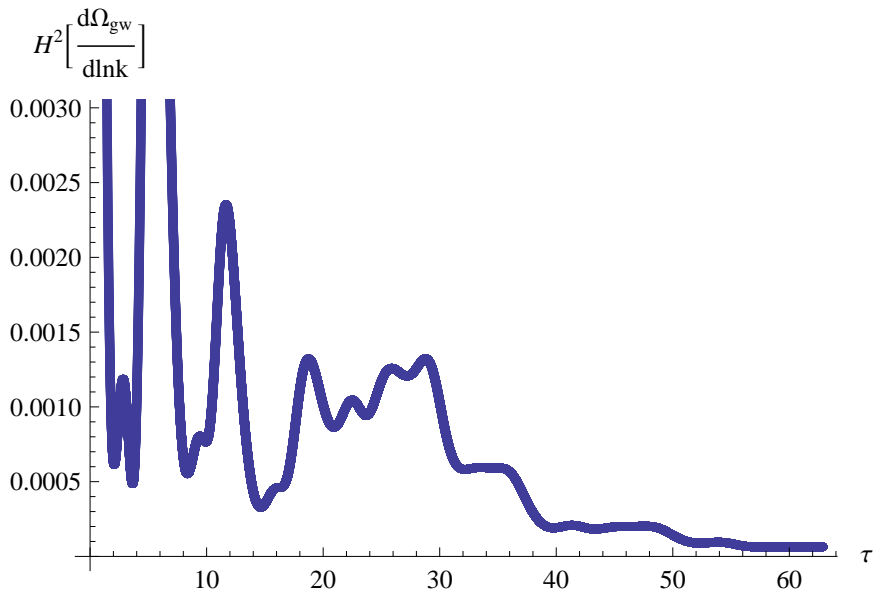,height=3.1in}}
\caption{\footnotesize Time evolution of $\; {\cal H}^2 
\times (\frac{d \;\;\;}{d \ln k} \, \Omega_{\rm gw}) \,$ 
for $\kappa = 2$ during the oscillatory regime.}
\label{rhoH}
\end{figure}

\newpage

\section{Physical Consequences}

The results of the previous Section allow us to make
the following remarks: 
\\ [5pt]
$\bullet \;$ The {\it existence of the enhancement effect} 
is confirmed by our analysis. In Section 4 we argued, on 
physical grounds, that the effect is associated with sign 
changes of the Hubble parameter. This is explicitly seen 
in Figures 13-14 where there is a synchronization among the 
strongest enhancement peaks of the observable (Fig. 14) and 
sign changes of the Hubble parameter (Fig. 13). Moreover, 
we see that the effect diminishes with time.
\\ [5pt]
$\bullet \;$ The {\it far super-horizon} modes left 
the horizon many e-foldings before criticality, their
mode functions are ``frozen'' thereafter, and these 
modes are not affected much from the presence of the 
oscillating regime.
\\ [5pt]
$\bullet \;$ The {\it near-horizon} modes show the 
enhancement due to resonances close to the oscillatory 
era frequency $\omega$. Notice that as $N_{\rm osc}$ 
increases the peak enhancement magnitude increases 
and shifts towards higher values of $\kappa$. Thus,
it is the {\it near/sub-horizon} modes that feel
the biggest enhancement.
\\ [5pt]
$\bullet \;$ The {\it far sub-horizon} modes show an 
ever-increasing ``tail'' with increasing $\kappa$
(Figs. 1-11). This enhancement is there and should be 
observed if very high frequency gravity waves become 
detectable in the future. It is present even when 
$N_{\rm osc} = 0$ (Fig. 1) and, hence, it has nothing 
to do with the existence or not of the oscillatory 
epoch. Furthermore, we can understand it without 
resorting to numerical results. When $\kappa \gg 1$ 
the variable (\ref{Mexpdimless}) ${\cal M}$ becomes 
essentially unity and the observable (\ref{detectorsdimless})
simplifies considerably:
\begin{equation}
\Big( \frac{{\cal H} \alpha}{\kappa} \Big)^2
\; \ll \; 1
\quad \Longrightarrow \quad
{\cal M} \; \approx \; 1
\quad \Longrightarrow \quad
\frac{d \;\;\;}{d \ln k} \, \Omega_{\rm gw}
\; \approx \;
\# \times \frac{\kappa^2}{\alpha^2}
\;\; . \label{detectorsfarsub}
\end{equation}
Therefore, at any fixed time $\tau$, the value of the
observable -- being proportional to $\kappa^2$ -- 
will follow a parabola as $\kappa$ increases; this is 
explicitly seen in Figures 1-11. At any fixed $\kappa$, 
the observable is proportional to $\alpha^{-2}$ and its 
value decreases accordingly as $\tau$ increases; this
is seen in Figure 14, albeit for $\kappa = 2$.
\\ [5pt]
$\bullet \;$ The {\it high-frequency ``tail''} discussed 
above will inevitably lead to ultraviolet divergences. 
Ultimately it is the correct ultraviolet theory of quantum 
gravity that will have to address the issue. Nonetheless, 
an interesting question to answer -- within the framework 
of ordinary perturbative quantum gravity -- is which 
counterterms would absorb the ultraviolet divergences of 
our observable. 

As a first step, we use (\ref{Mexpdimless}) to expand 
(\ref{detectorsdimless}) in powers of 
$( \frac{{\cal H} \alpha}{\kappa} )^2$ until we reach
ultraviolet convergence. To make the connection with the 
available counterterms more direct, we also convert -- 
using (\ref{detectors1}) -- from the ratio $\Omega_{\rm gw}$ 
to the excess energy density $\Delta \rho$. The result is:
\begin{equation}
\frac{d \;\;\;}{d \ln k} \, \Delta \rho (t, k) = 
\frac{\omega^4}{8 \pi^2} \left\{
{\cal H}^2 \Big( \frac{\kappa}{\alpha} \Big)^2 +
\Big[ \frac32 \, \epsilon - \frac34 \, \epsilon^2 +
\frac{\dot{\epsilon}}{2 H} \Big] {\cal H}^4 + 
O \Big( \frac{{\cal H}^6 \alpha^2}{\kappa^2} \Big)
\right\}
\label{Deltarhoexp}
\end{equation}
To make (\ref{Deltarhoexp}) ultraviolet finite, two
subtractions are needed: one to renormalize the 
$\kappa^2$ term and one to renormalize the constant
term. As we shall see, the two counterterms which
absorb the divergences are, respectively:
\begin{equation}
\Delta {\cal L}_1 \; = \;
g_1 \, R \, \sqrt{-g}
\qquad , \qquad
\Delta {\cal L}_2 \; = \;
g_2 \, R^2 \, \sqrt{-g}
\;\; . \label{counterterms}
\end{equation}
These counterterms induce the following stress-energy
tensor contributions:
\begin{eqnarray}
\Delta T^1_{\mu\nu} &\!\! = \!\!&
2 g_1 \Big[ R_{\mu\nu} - \frac12 \, g_{\mu\nu} R \Big]
\;\; , \label{DeltaTmn1} \\
\Delta T^2_{\mu\nu} &\!\! = \!\!&
2 g_2 \Big[ 2 R_{\mu\nu} - \frac12 \, g_{\mu\nu} R +
2 \Big( g_{\mu\nu} \square - D_{\mu} D_{\nu} \Big)
\Big] R
\;\; . \label{DeltaTmn2}
\end{eqnarray}
We are interested in the energy density component of
the stress-energy tensor for cosmologically relevant
$(FRW)$ spacetimes; in that case:
\begin{eqnarray}
FRW 
\quad \Longrightarrow \quad
\Delta T^1_{00} &\!\! = \!\!&
6 \, g_1 \, \omega^2 \, {\cal H}^2
\;\; , \label{DeltaT001} \\
FRW 
\quad \Longrightarrow \quad
\Delta T^2_{\mu\nu} &\!\! = \!\!&
- 144 \, g_2 \; \omega^4 
\Big[ \frac32 \, \epsilon - \frac34 \, \epsilon^2 +
\frac{\dot{\epsilon}}{2 H} \Big] {\cal H}^4 
\;\; , \label{DeltaT002}
\end{eqnarray}
and our assertion is established: $\Delta T^1_{00}$ --
given by (\ref{DeltaT001}) -- absorbs the quadratically
diverging order ${\cal H}^2$ term in (\ref{Deltarhoexp}), 
while $\Delta T^2_{00}$ -- given by (\ref{DeltaT002}) -- 
absorbs the order ${\cal H}^4$ term in (\ref{Deltarhoexp}) 
which diverges logarithmically.
\footnote{The value of $g_2$ needed to subtract the order
${\cal H}^4$ term in (\ref{Deltarhoexp}) agrees with that 
first found in 1974 by `t Hooft and Veltman \cite{thooft}.}
\\ [5pt]
$\bullet \;$ The {\it present value of the enhancement}
can be straightforwardly computed:
\begin{eqnarray}
\left( \frac{d \;\;\;}{d \ln k} \, 
\Omega_{\rm gw} \right)_{\rm now}
&\!\! \approx \!\!&
\Big( \frac{a_{\rm matter}}{a_{\rm now}} \Big) 
\left( \frac{d \;\;\;}{d \ln k} \, 
\Omega_{\rm gw} \right)_{\rm matter}
\label{deterctorsnow1} \\
&\!\! \approx \!\!&
\Big( \frac{a_{\rm matter}}{a_{\rm now}} \Big) 
\left( \frac{d \;\;\;}{d \ln k} \, 
\Omega_{\rm gw} \right)_{\rm osc}
G \omega^2
\label{detectorsnow2} \\
&\!\! \approx \!\!&
0.3 \times 10^{-3} 
\left( \frac{d \;\;\;}{d \ln k} \, 
\Omega_{\rm gw} \right)_{\rm osc}
4 \times 10^{-12}
\;\; . \label{detectorsnow3}
\end{eqnarray}
The value of the observable in the oscillatory regime,
for given $N_{\rm osc}$ and $\kappa$, can be found in 
Figures 1-11. 
\footnote{As mentioned in Section 6, the displays of
the observable $\left( \frac{d \;\;\;}{d \ln k} \, 
\Omega_{\rm gw} \right)_{\rm osc}$ in all Figures
therein lack an overall factor of $G \omega^2$.}
The remaining factor of about $10^{-15} $ in 
(\ref{detectorsnow3}) makes the enhancement effect 
very small and presently unobservable.
The passage from (\ref{detectorsnow2}) to
(\ref{detectorsnow3}) is valid because after their
entrance to the radiation era the gravitational waves 
behave like any other kind of radiation, and because 
during the radiation regime -- unlike the matter regime 
-- the product ${\cal H}^2 \alpha^4$ appearing in 
(\ref{detectorsdimless}) is constant.
\\ [5pt]
$\bullet \;$ The {\it present frequency} of the enhanced 
waves is given by:
\begin{equation}
f_{\rm now} \; = \;
\frac{k}{2 \pi \, a_{\rm now}} 
\; = \;
\frac{\omega \kappa}{2 \pi} \,
\Big( \frac{a_{\rm cr}}{a_{\rm now}} \Big) 
\; \ltwid \; 
10^9 H\!z \times e^{- \frac12 \Delta N}
\; \ltwid \; 
10^9 H\!z \
\;\; . \label{fnow}
\end{equation}
For the estimate (\ref{fnow}) we used the 
{\it near-horizon} value $\kappa \approx 1$ 
as well as \cite{romania}:
\begin{eqnarray}
\Big( \frac{a_{\rm cr}}{a_{\rm now}} \Big) 
&\!\! \ltwid \!\!& 
e^{-63 - \frac12 \Delta N}
\; \approx \;
10^{-28} \times e^{- \frac12 \Delta N}
\;\; , \label{aratiovalue} \\
\omega &\!\! \ltwid \!\!& 10^{55} H_{\rm now}
\; \approx \; 3.2 \times 10^{37} H\!z
\;\; , \label{omegavalue}
\end{eqnarray}
where $\Delta N$ is the number of oscillatory 
e-foldings which we expect to be small.

\section{Epilogue}

From Figure \ref{rhoH} it is evident that the signal is
peaked at a narrow band of very high frequencies and is 
negligible at significantly different frequencies. It 
would be challenging to detect gravitational radiation
at such high frequencies but detectors in that range 
have been proposed \cite{gravwaves}. As noted in the 
text, the phase of oscillations does not affect modes 
which experienced first horizon crossing more than a 
few e-foldings before the end of inflation. The wavelength 
of our effect is $\, \lambda = f^{-1} \gtwid 0.3 m$, 
whereas the smallest scale feature which is currently 
observed in the cosmic microwave radiation is about 
$10^{22} m$ \cite{gravwaves}! Our model does not change 
either how matter couples to gravity or the propagation 
of linearized gravitons, so it has no effect on the 
spin-down rate of the binary pulsars. The gravity waves 
we predict will certainly distort how pulsar light 
propagates, but the short wavelength again seems to 
preclude a detectable effect. LIGO is not sensitive 
above frequencies of $\, 7000 H\!z$, which is far too 
low. The situation is even worse with LISA's high 
frequency cutoff of $\, 0.1 H\!z$ \cite{gravwaves}.

\newpage

\centerline{\bf Acknowledgements}
We should like to thank Neil Cornish for conversations.
This work was partially supported by the European 
Union grant FP-7-REGPOT-2008-1-CreteHEPCosmo-228644, 
by the NSF grant PHY-0855021, and by the Institute for 
Fundamental Theory at the University of Florida.

\end{document}